\documentstyle{lamuphys}
\makeatletter
\let\chapter\hid@chapter
\makeatother
\begin{document}
\pagenumbering{arabic}
\title{Infrared imaging and off-nuclear spectroscopy of quasar hosts}

\author{Marek J.\,Kukula\inst{1}, James S.\,Dunlop\inst{1}, David
H.\,Hughes\inst{1}, Geoff\,Taylor\inst{2}\\ and Todd\,Boroson\inst{3}}

\institute{Institute for Astronomy, University of Edinburgh,
Royal Observatory, Edinburgh EH9 3HJ, UK
\and
Astrophysics Group, Liverpool John Moores University,
Byrom St, Liverpool L3 3AF, UK
\and
NOAO, PO Box 26732, Tucson, AZ 85726-6732, USA}

\authorrunning{Kukula {\it et al.}}
\maketitle

\begin{abstract}

We present the results of two complementary ground-based programmes to
determine the host galaxy properties of radio-quiet and radio-loud
quasars and to compare them with those of radio galaxies. Both
infrared images and optical off-nuclear spectra were obtained and we
discuss the various strategies used to separate the quasar-related
emission from that of the underlying galaxy.  However, the key feature
of this project is the use of carefully matched samples, which ensure
that the data for different types of object are directly comparable.

\end{abstract}

\section{Introduction}

The paper briefly describes the results of a continuing long-term
project to study the host galaxies of powerful AGN. The aim of the
project is two-fold: to test the scheme for radio-loud quasars (RLQs)
and radio galaxies (RGs) which attemps to unify the two types of
object via orientation effects, and to investigate the extent to which
the host galaxy influences the radio properties of the AGN by
comparing the hosts of radio-loud and radio-quiet quasars (RQQs).

Using ground--based observations we have approached the question of
host galaxy properties from two independent directions: near-infrared
($K$-band) imaging, to determine the host morphologies and
luminosities, and off-nuclear optical spectroscopy to investigate
their star-formation histories. A more detailed description of the
near-infrared imaging can be found in Dunlop {\it et al.} (1993) and
Taylor {\it et al.} (1996).

\section{Sample selection}

In the past, comparative studies of quasars and radio galaxies have
often been hampered by the use of poorly matched samples, sometimes using
wildly differing selection criteria. A key feature of the current
project was the selection of three {\it carefully matched samples}
of RQQs, RLQs and powerful (FRII) radio galaxies.

In order to ensure that the samples were directly comparable with each
other the RQQs and RLQs were selected to have identical distributions
in the $V-z$ plane. Meanwhile the RG sample was chosen to match the
radio luminosity $-$ redshift ($L_{5GHz} - z$) and spectral index -
redshift ($\alpha - z$) distributions of the RLQs.  The objects are
all of relatively low redshift and cover a narrow range in $z$ ($0.1 <
z < 0.3$). Both of the quasar samples were drawn largely from the
(optically-selected) Bright Quasar Survey (Schmidt \& Green 1983) and
consist of objects at the fainter end of the quasar luminosity
function ($-26 < M_{B} < -23$).

\section{Near-IR imaging with UKIRT}

There are several advantages to working at near-infrared
wavelengths. Quasars are, by definition, heavily nuclear-dominated
objects in the optical, making galaxy magnitude and morphology
determination extremely sensitive to the estimated strength of the
core component and the reliability of the adopted form of the
point-spread function. Quasars are relatively blue objects ($f_{\nu}
\simeq$~const) whereas the luminosity of the host galaxy is expected
to peak at near-infrared wavelengths (Sanders {\it et al.} 1989),
making the near-infrared the waveband of choice for minimising the
nuclear:host ratio. Working in the near-infrared also helps to avoid
contamination of the images by strong emission lines and/or light from
regions of enhanced starformation, both of which could mask the true
nature of the underlying galaxy. Finally, the high sky background in
the infrared - a major drawback in many ways - does at least mean that
the signal:noise ratio of an image is not compromised by subdividing
the integration into sufficiently small sub-integrations to avoid
saturation of the quasar nucleus, a point of some importance when
attempting to determine the correct form of the PSF.

\subsection{Observations and modelling}

The observations were made in $K$-band (2.2$\mu$m) using the $62
\times 58$ array IRCAM on the United Kingdom Infrared Telescope
(UKIRT). A library of $\sim 100$ standard star images was compiled
from which the PSF most suitable for a particular quasar image could
subsequently be selected.  The procedure adopted for the data
reduction is described in detail by Taylor {\it et al.} (1996).

In our first attempt to analyse the quasar images a simple PSF
subtraction was used (Dunlop {\it et al.} 1993) but this gave rise to
a number of problems; most significantly there was an inevitable
oversubtraction of galaxy light in the centre of the image.  The
adopted solution was to use two-dimensional modelling of the surface
brightness distribution in order to properly decouple the shape of the
host galaxy from that of the PSF (Taylor {\it et al.}  1996). This
allows one to extrapolate smoothly into the central regions of the
galaxy and thus estimate the luminosity of each host in a
self-consistent manner.

The modelling algorithm fits five parameters to the data: nuclear
luminosity, host galaxy luminosity, galaxy scale-length, galaxy
position angle and axial ratio. In order to determine the
morphological type of the host galaxy one further parameter is
required: the index $\beta$, which describes the form of the galaxy
luminosity profile ($\mu(r)/\mu_{0} = exp((-r/r_{0})^{\beta})$). For
the present study we decided to consider only the alternatives of
$\beta = 1$ (an exponential disc) and $\beta = 1/4$ (a de Vaucouleurs
law), and thus to confine our morphological investigation to
determining whether a given host galaxy is dominated by a disc or a
spheroidal component. For each quasar the fitting procedure was
carried out twice, once with $\beta = 1$ and again with $\beta = 1/4$,
and the two model fits were then compared with the original image to
determine which was the most successful.

In view of the large number of free parameters involved in the model
fitting procedure, as well as the notorious sensitivity of such
methods to uncertainties in the adopted form of the PSF, the process
was subjected to rigorous testing in order to establish the degree of
reliability of the fits under a wide range of starting conditions.  To
this end a series of synthetic quasar$+$host combinations were
constructed and convolved with a range of PSFs. The accuracy with
which the modelling algorithm was able to recover the `true'
parameters of the artificial galaxies could then be measured in a
self-consistent fashion. These tests show that the host galaxy
parameters derived from the model fits are typically accurate to
within 10\%. 

However, as expected, the uncertainty increases with the nuclear:host
ratio of the quasar and this also has a very strong effect on the
degree of confidence with which an exponential disc profile can be
distinguished from a de Vaucouleurs law. We decided to adopt a
pessimistic approach and to reject as unreliable the morphological
classification of objects for which the nuclear:host ratio exceeded a
very conservative limit. Effectively, this means that for $z\simeq
0.1$ objects with $L_{nuc}:L_{host}>10$ are excluded; for $z\simeq
0.3$ the limit becomes an even more stringent
$L_{nuc}:L_{host}>5$. Fortunately, the low nuclear:host ratio of
quasars at near infrared wavelengths means that the majority (31/40)
of the objects in our samples survive this selection procedure, and in
these cases we are confident that the morphological preference
displayed by the fitting algorithm is both valid and meaningful.  This
would not have been the case in the optical, where typical values of
$L_{nuc}:L_{host}$ exceed 10 (in $B$-band; see Taylor \& Dunlop 1997).

\subsection{Results}

The principal results to emerge from this study can be summarized as
follows:

(i) RGs, and the hosts of both RLQs and RQQs are all {\bf luminous
galaxies} with $L\geq L^{\star}$ at $K$ ($<M_{K}> \simeq -26$).

(ii) RGs and the hosts of RLQs and RQQs are all {\bf large galaxies}
with a half-light radius ({\it ie} the radius containing half of the
total galaxy luminosity) $r_{1/2} \geq 10$~kpc.

(iii) The basic parameters of the host galaxies are no different from
those of other comparably large and luminous galaxies. In particular
the hosts of all three types of AGN display a $\mu_{1/2} - r_{1/2}$
relation which is identical in both slope and normalisation to that
displayed by brightest cluster galaxies - objects which are thought to
be the product of successive merger events. This suggests that,
regardless of their current interaction status, the host galaxies of
powerful AGN have all experienced merger events in the past.

(iv) Essentially all of the RGs and RLQ hosts are best described by a
de Vaucouleurs law, consistent with unification of powerful radio-loud
AGN via orientation. Thus it appears that an elliptical host galaxy is
necessary for an active galaxy to produce a radio luminosity in excess
of $L_{5GHz}\simeq 10^{24}$WHz$^{-1}$sr$^{-1}$.

(v) Slightly more than half of the RQQs appear to lie in galaxies
which are dominated by an exponential disc. Those RQQs which have
elliptical hosts are in general more luminous than those which reside
in discs. A significant fraction of the RQQ population may at least
be capable of producing powerful radio emission.

(vi) The majority of the radio galaxies in our sample contain
additional nuclear flux at $K$ in excess of that expected from the
best fitting $r^{1/4}$-law model. These unresolved nuclear components
may simply be indicative of central cusps in their starlight, but
their colours and magnitudes are consistent with dust-reddened
quasars (Taylor \& Dunlop 1997).

\subsection{Comparison with HST results}

In general the findings of our ground-based imaging programme are in
good agreement with those of recent optical HST studies ({\it eg}
Hutchings {\it et al.} 1994, Hutchings \& Morris 1995, Disney {\it et
al.} 1995). The red colours and large scale-lengths of the hosts as
determined from our ground-based data almost certainly explain the
failure of earlier HST programmes to detect some of these galaxies in
the optical ({\it eg} Bahcall, Kirhakos \& Schneider 1995). Subsequent
re-analysis of these HST images has revealed that large, luminous host
galaxies are indeed present (McLeod \& Rieke 1995, Bahcall, Kirhakos,
Saxe \& Schneider 1997).

\section{Off-nuclear optical spectroscopy}

A completely independent way to characterise the host galaxies of AGN
is via analysis and classification of their stellar populations. The
aim of the observations described in this section is to obtain high
signal-to-noise spectra of the quasar hosts which could then be used
to determine the composition, age and evolutionary history of their
stellar components.  

Previous attempts to take optical spectra of quasar `fuzz' were
severely hampered by scattered light from the quasar itself which
effectively swamped the starlight from the surrounding galaxy and
prevented any meaningful analysis from being carried out. However,
the deep near-infrared images of our quasar samples presented us with
a unique opportunity to circumvent this problem: armed with knowledge
of the extent and orientation of the host galaxy on the sky we were
able to choose a slit position which was far enough from the nucleus
to avoid the worst excesses of scattered quasar light, but which
simultaneously maximised the amount of galaxy light falling onto the
slit.

\begin{figure}
\vspace{8.0cm}
\includegraphics{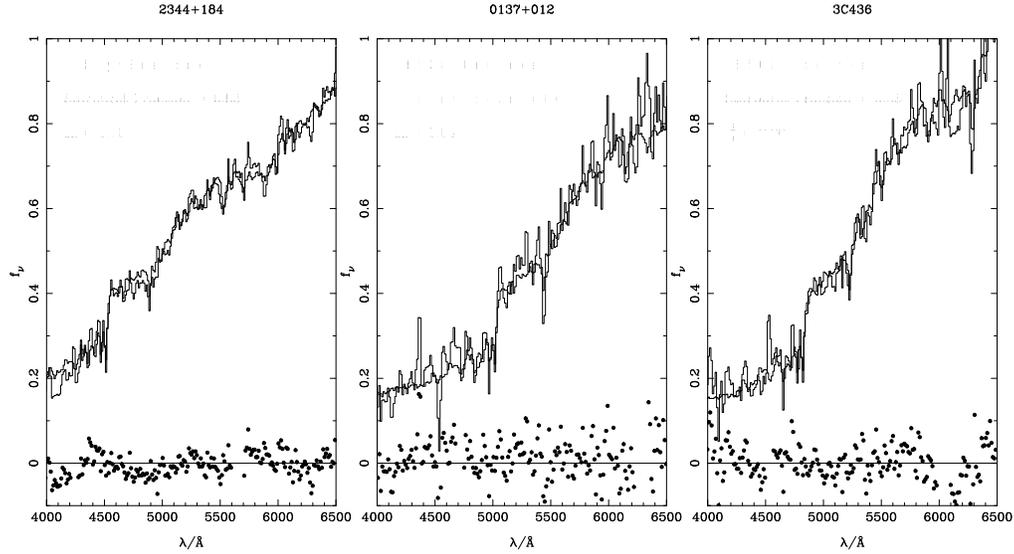}

\caption{Off-nuclear spectra of three active galaxies: the RQQ
2344$+$184 ($z = 0.137$), the RLQ 0137$+$012 ($z = 0.258$) and the radio
galaxy 3C436 ($z=0.215$). Each spectrum is best described by the
combination of an old burst model and a very blue component (probably
scattered quasar light). The observed spectrum is shown as a heavy
line, the model by a thin line and the residuals by dots.}

\end{figure}

\subsection{Observations}

Initial observations were carried out on 10 objects using the Mayall
4-m telescope at Kitt Peak and covering a wavelength range of
3500-7500\AA. With the slit positioned $5''$ from the nucleus,
starlight was easily detected in all 10 objects and the spectra were
of sufficient quality to allow us to fit spectrophotometric models to
the stellar populations.

Subsequent observations were carried out on the 4.2-m William Herschel
Telescope (WHT) on La Palma. The availability of the ISIS double-beam
spectrograph on the WHT enabled us to extend our wavelength range into
the red down to $\sim9000$\AA~- the extra wavelength coverage being
particularly useful for constraining models of galaxy
spectrophotometric evolution. To date twenty five of our objects have
been observed.

\subsection{Initial analysis}

For the purposes of this early analysis we have attempted to fit a
simple `burst' model (Guiderdoni \& Rocca-Volmerange 1987) to the
off-nuclear spectrum, varying the age of the model to obtain the best
fit. The burst model assumes that all starformation occurs in a single
burst of activity lasting $\sim$1~Gyr, and that the stellar population
evolves passively thereafter. We are currently working towards
applying more sophisticated and realistic models, but we note that the
fits obtained using this very simple scenario are surprisingly good
(Figure~1).

Whilst longwards of the 4000\AA~ break the spectra are clearly
dominated by starlight, at shorter wavelengths an additional blue
component becomes prominent in some of the off-nuclear quasar spectra.
This leads to poor fits from the burst model and to extremely young
ages for the stellar population.

Since the presence of this blue component is often accompanied by the
appearance of broad emission-lines, particularly H$\beta$, we
conjectured that it was probably the result of residual scattered
quasar light which, although relatively low-level, was still
sufficient to dominate the combined spectrum at $\lambda_{rest} <
4000$\AA.  In order to test this theory we took a nuclear spectrum of
the RQQ 0054$+$144 and scaled it to match the height of the broad
H$\beta$ feature in the off-nuclear spectrum of the same
object. Subtraction of the scaled nuclear spectrum caused a marked
improvement in the quality of the resulting fit and, as expected, a
substantial increase in the age of the best fitting model (from 5 to
13~Gyrs).

We therefore decided to carry out a two-component fit to the
off-nuclear spectra, using a combination of a burst model and a
flat-$f_{\nu}$ component to simulate scattered quasar light, and
allowing the age of the burst and the amplitude of the flat component
to vary freely. For the sake of consistency this procedure was used on
all the objects in our sample, including the radio galaxies and those
quasars which appeared to lack a strong blue component. This approach
appears to have been vindicated by the fact that in cases where a blue
component was not obviously present in the spectrum the modelling
algorithm invariably achieved a good fit without resorting to the
addition of a strong `scattered quasar' component. As a final check,
we applied the same procedure to spectra of M32, a dwarf elliptical
companion of the Andromeda galaxy, and M33, a nearby late-type
spiral. In both cases a good fit was obtained without recourse to a
flat-$f_{\nu}$ component, and sensible ages were obtained from the
models (old for the dwarf elliptical and relatively young for the
spiral).

\begin{figure}
\vspace{18.0cm}
\includegraphics{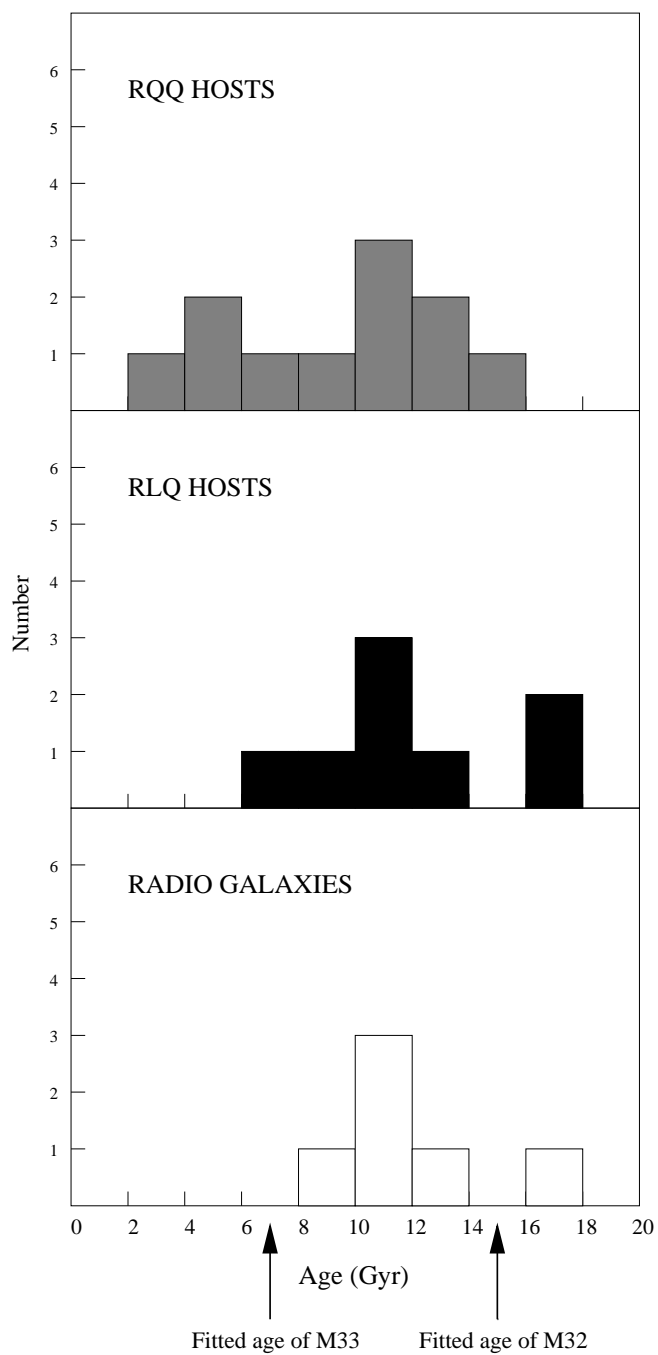}

\caption{Age distributions of the stellar populations in our three
matched samples. Note that the RQQ sample contains a significant
proportion of young, blue hosts whereas the RLQ and RG samples have
distributions which are indistinguishable from each other and consist
of older, redder galaxies. Also indicated in this figure are our fits
to M33 (a late-type spiral) and M32 (a dwarf elliptical).}

\end{figure}

\subsection{Preliminary results}

The galaxy ages obtained from the best fitting models are shown in
Figure~2 along with the derived ages for M32 and M33. The histograms
for the RLQs and RGs are statistically indistinguishable and the host
galaxies are generally rather old, red systems, consistent with
unification of the two types of AGN.  The histogram for the RQQs shows
a prominent tail of younger, bluer galaxies - three galaxies have ages
$<6$~Gyr even after the removal of any scattered quasar light. A
comparison with our near-infrared images shows that all of these
`young', blue galaxies appear to have close companions and/or display
distorted morphologies, implying that they are currently (or have
recently been) involved in interactions or mergers.

However, the general trend is that the hosts of all three types of AGN
are dominated by an old stellar population. In the RG and RLQ samples
80\% and 75\% respectively of the host galaxies have ages $\geq
11$~Gyr, whilst for the RQQ sample the proportion with ages $\geq
11$~Gyr is still 50\% (with the current, rather crude level of
analysis, distinguishing between ages greater than 11~Gyr is a highly
model-dependent affair). Other than the fact that the youngest RQQ
hosts appear to be interacting galaxies there is no obvious
correlation between the age of the hosts and their morphological type.

\section{Summary}

This has proved to be a very fruitful project. Many interesting (and
some unexpected) results have emerged from the near-infrared imaging
study and, although the off-nuclear spectroscopy is still very much a
work in progress, the fact that we have been able to isolate starlight
in all of the spectra taken so far is a very encouraging result. 

A consistent picture is emerging from the data. It appears that RLQ
hosts and RGs are indeed the same type of galaxy - large luminous
spheroidal systems with old, red stellar populations - consistent with
the unified scheme. The hosts of RQQs are also large and luminous, and
can be either disc-dominated or spheroidal systems.  There seems to be
a tendency for the most luminous RQQs to occur in elliptical
hosts. Ages of the RQQ hosts cover a wide range and the bluest galaxies
all seem to be undergoing interactions.

In the immediate future we have been awarded 34 orbits on the HST to
observe our three AGN samples in $R$-band. Not only will these images
enable us to determine the optical morpholgies of the hosts with a
level of detail which is impossible from the ground, but by providing
us with reliable optical luminosities for the host galaxies they will
enable us to bridge the gap between our two ground-based datasets,
allowing us to calculate $R-K$ colours for the galaxies and thus to
test whether a particular spectrophotometric model can explain the
shape of a galaxy spectrum from optical through to near-infrared
wavelengths.

\end{document}